\documentclass{eas}
\usepackage{graphicx}
\TitreGlobal{The Future Astronuclear Physics}
\begin{document}

\title{Core-Collapse Supernovae Induced by \\ Anisotropic Neutrino Radiation} 
\runningtitle{Motizuki \etal: Core-Collapse Supernovae with Anisotropic $\nu$ Radiation}
\author{Yuko Motizuki}\address{RIKEN, Hirosawa 2-1, Wako 351-0198 Japan}
\author{Hideki Madokoro}\sameaddress{1}
\author{Tetsuya Shimizu}\sameaddress{1}
\begin{abstract}
We demonstrate the important role of anisotropic neutrino radiation on 
the mechanism of core-collapse supernova explosions.
Through a new parameter study with a fixed radiation field of neutrinos,
we show that prolate explosions caused by globally anisotropic 
neutrino radiation is the most effective mechanism of increasing 
the explosion energy when the total neutrino luminosity is given.
This is suggestive of the fact that the expanding materials of 
SN~1987A has a prolate geometry.
\end{abstract}
\maketitle

\newcommand{\apj}{Astrophys. J.}
\newcommand{\aap}{Astron. Astrophys.}
\newcommand{\msolar}{$M_{\odot}$\,}

\section{Introduction}
In spite of every effort over more than 30 years since the first work of Colgate \& White (1966), 
the mechanism of core-collapse supernovae is still under debate.
In the so-called ``delayed explosion'' scenario, 
which is believed to be responsible
in the mechanism of core-collapse supernovae, 
the supernova shock is once stagnated around $\sim$150-200 km above the neutrinosphere.
Then neutrinos heat the matters behind the stalled shock, and finally help the shock wave
to revive or move upward, resulting in an explosion.
So far, no ``successful'' supernova simulations have been reported
except work by Wilson and Mayle (1988, 1993).
Here a ``successful" supernova explosion must explain the following observed facts:
1) the explosion energy (observed in SN~1987A): 1.5 $\pm \, 0.5 \times 10^{51}$ ergs,
2) asymmetry observed by spectropolarimetry measurements in SN~1987A, SN~1993J, and other 
several supernovae,
3) remnant neutron star mass of 1.4 \msolar, and
4) amount of explosive nucleosynthesis of $^{56}$Ni: $\sim$ 0.07 \msolar 
in the case of SN~1987A.

It has been theoretically demonstrated that simulations assuming spherical
symmetry fail to produce robust explosions even if general relativistic Boltzmann
neutrino transport is taken into account (Liebend\"orfer \etal~2001).
Therefore together with the above-mentioned observed asymmetries in the explosions 2), 
we call for multi-dimensional hydrodynamical simulations.
At this time, such simulations have been 
performed for two-dimensional (2D) models (Miller \etal~1993; Herant \etal~1994; 
Burrows \etal~1995; Janka \& M\"{u}ller 1996;
Mezzacappa \etal~1998; Fryer \& Heger 2000; Shimizu \etal~2001;
Buras \etal~2003) and three-dimensional models (Shimizu \etal~1993; Fryer \& Warren~2002).
In these simulations, special attention is paid
to the role of convection either near the surface of a nascent neutron star or
in neutrino-heated regions above the neutrinosphere.  
It has been argued that
large-scale mixing, caused by convection and convective overturn around the
neutrino-heated region, increases the average entropy and therefore the explosion 
energy, and this can trigger a
successful explosion (e.g., Herant \etal~1994; Janka \& M\"{u}ller 1996).
It is noted, however, that almost all hydrodynamical simulations so far have been 
performed with spherically symmetric neutrino radiation field. 
In this article, we show that {\em anisotropic} neutrino radiation,
or {\em locally intense} neutrino heating, can be an alternative 
mechanism to revive the shock wave and 
to lead to a successful explosion other than the so far suggested 
``convective trigger".
\\ \hspace*{1.3em}
Shimizu and coworkers (Shimizu \etal~1994; Shimizu \etal~2001) 
first proposed that the anisotropic neutrino radiation
should play a crucial role in the explosion mechanism 
and carefully investigated the effects of anisotropic neutrino radiation on the 
explosion energy.
They performed 2D hydrodynamical simulations and found that 
only a few percent of enhancement in the neutrino flux along the axis of symmetry
is sufficient to increase the explosion energy by a large factor, 
and that this effect saturates around a certain degree of anisotropy.
\\ \hspace*{1.3em}
The origin of anisotropic neutrino radiation is considered as follows.
Because supernova progenitors such as OB stars are generally observed
to be fast rotators ($\sim 200$ km s$^{-1}$ at the surface, $P\sim 1$ day;
see, e.g., Fukuda 1982), the
resulting proto-neutron stars can have a large amount of angular momentum
after the gravitational collapse.  
Centrifugal force then deforms the rotating core into an oblate form.  
This will cause asymmetric neutrino radiation, in
which the flux along the pole is enhanced over that on the equatorial plane.
Anisotropy in neutrino radiation may also be originated from the convection inside
the proto-neutron star, or asymmetric mass accretion onto the neutron star.
Recently, Kotake \etal~(2003) performed 2D simulations of the
rotational collapse of a supernova core and found that
the rotation can actually produce stronger neutrino radiation
in the direction of the rotational axis than that on the equator.
\\ \hspace*{1.3em}
Neutrino flux can also fluctuate 
with angle and time as suggested by Burrows \etal~(1995), 
due to gravitational oscillation of the neutrinosphere.
In this article we carry out 2D
hydrodynamical simulations with a fixed radiation field
of neutrinos, to investigate what kind of neutrino
radiation is favorable for a successful explosion.
We briefly demonstrate that
globally anisotropic (prolate) neutrino radiation is the most effective way
of increasing the explosion energy compared with spherically symmetric and 
fluctuated neutrino radiation.
Details are found in Madokoro \etal~(2003, 2004).

\section{General Features of an Explosion Induced by Anisotropic $\nu$ Radiation}

In this section we describe our simplifications to the problem 
and general features of an explosion induced by anisotropic neutrino radiation. 
We perform 2D simulations by solving hydrodynamical equations in the
spherical coordinate ($r,\theta$).  
A generalized Roe's method is used
to solve the general equations of motion.  
In order to set the initial condition, 
we use the fact that, in the delayed mechanism, the shock is stalled 
at a radius of a few hundred kilometers and stays for a few hundred milliseconds.
We reproduce the stalled shock wave at the radius of 200 km 
by solving stationary 
hydrodynamic equations with assuming spherical symmetry, and start
calculations when a stalled shock wave is formed.
It is a good approximation to exploit the stationary hydrodynamic solution
as an initial model.
Here the radius of the proto-neutron star is fixed to be 50 km
and the computational region ranges from 50 km to 10000 km in radius 
from the center of the proto-neutron star.  
The details of our numerical technique, the adopted equation of state,
and the initial conditions are described in the previous papers
(Shimizu \etal~2001; Madokoro \etal~2003).
It is noted here that we have improved the numerical code in these papers 
to avoid a numerical error near the pole.
Note that this numerical error was not serious and minor
for the investigation of the explosion energy, but may affect the results
of nucleosynthesis.

The local neutrino flux seen by an observer positioned 
at a distance $r$ far from the neutrinosphere
and at the angle $\theta$ from the axis of symmetry
is assumed as
\begin{equation}
  l_\nu(r,\theta) = \frac{7}{16}\sigma T_{\nu}^{4} c_{1}
  \left(1+c_{2}\cos^{2}\theta\right)\frac{1}{r^{2}},
  \label{eqn:nuflux}
\end{equation}

\noindent
where $\sigma$ is the Stefan-Boltzmann constant
and $T_{\nu}$ is the neutrino temperature of assumed blackbody radiation.
In our simulations $T_{\nu}$ is taken to be constant 
and hence the total neutrino luminosity, $L_{\nu}$, which is obtained 
by integrating Eq.~(\ref{eqn:nuflux}) over the solid angle,
is also constant in time.
Recent calculations (e.g., Fryer \& Heger 2000) have suggested that
the decay timescale of the neutrino luminosity is about 500 ms,
and hence we stop our calculations with constant $L_{\nu}$ at t=500 ms 
after the shock stagnation.
The above simplifications are enough for our purpose
(see Madokoro \etal~2003 for details).

\begin{figure}[tb]
\begin{center}
\includegraphics[scale=0.7]{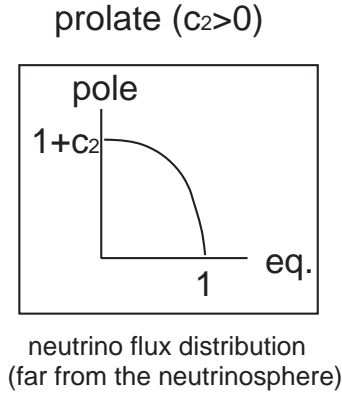}
\end{center}
\caption{
A schematic picture that shows the neutrino flux distribution 
seen from each respective observer far from the neutrinosphere.
}
\end{figure}

In Eq.~(\ref{eqn:nuflux}),
the parameter $c_{2}$ represents the magnitude of anisotropy in
the neutrino radiation.  
One can easily confirm that the neutrino fluxes in the polar and equatorial 
directions, $l_{z} \equiv l_{\nu}(r, \theta=0)$ and 
$l_{x} \equiv l_{\nu}(r, \theta=\pi/2)$, are proportional to
$c_{1}(1+c_{2})$ and $c_{1}$, respectively.  
The degree of anisotropy
seen from each respective observer is then given by
\begin{equation}
  [\frac{l_{z}}{l_{x}}]_{obs} = 1 + c_{2}.
  \label{eqn:anisotropy}
\end{equation}
This is schematically illustrated in Fig.~1.
The value of $c_{1}$ in Eq.~(\ref{eqn:nuflux}) is
calculated from given $c_{2}$ so as to adjust the total neutrino
luminosity $L_{\nu}$ to that in the spherical model at the same $T_{\nu}$.

\begin{figure}[tb]
\begin{center}
\includegraphics[scale=0.52]{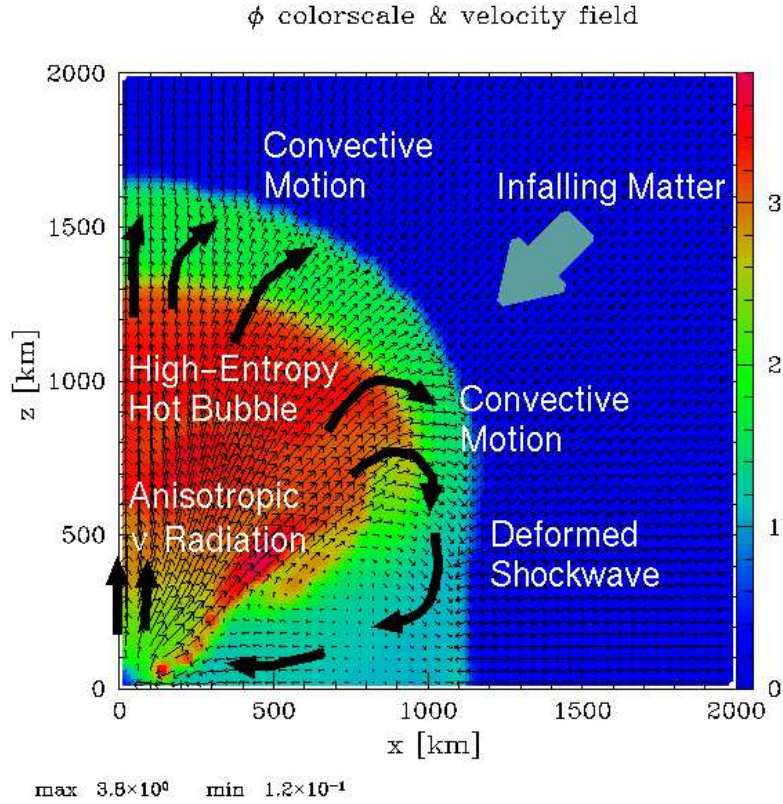}
\end{center}
\caption{
Dimensionless entropy contour map with the velocity field
for the model of $[l_{z}/l_{x}]_{obs}$ = 1.1 and 
$T_\nu$ = 4.7 MeV at t = 160 ms after the shock stall.
Note that the shock wave itself is deformed.
}
\label{fg:figure}
\end{figure}

Figure~2 depicts the dimensionless entropy contour and the velocity field
to explain the general feature of an explosion induced by anisotropic neutrino radiation.
As described above, a stronger neutrino radiation is assumed to be emitted in the 
direction of the symmetric (polar) axis
of a proto-neutron star than in the equatorial direction.
The matter surrounding the neutron star is first heated along the pole,
and a high-entropy region is formed there.
Since there is a gravitational pull by the central neutron star, the high-entropy
bubble moves upward as a result of buoyancy.
The upward velocity field can be clearly seen in Fig.~2.
The high pressure of the heated matter then pushes up the shock front, which deviates
from spherical symmetry.
This asymmetric, jetlike explosion is a common feature as a consequence of anisotropic 
neutrino radiation.
It has been found that the hot bubble evolves faster with increasing
the anisotropy in the neutrino radiation field.
The large mushroom-like structure of the hot bubble is produced, as seen in 
Fig.~2, as a result of the global circular motion behind the shock front.
The circulation is global and clockwise in the figure, and its cycle is only a 
few rotations per explosion. 
The distortion of the shock wave continues and not diminished in the course
of the shock propagation.

\section{Explosion Energy Increase Triggered by Anisotropic $\nu$ Radiation}

\begin{figure}[tb]
\begin{center}
\includegraphics[scale=0.42]{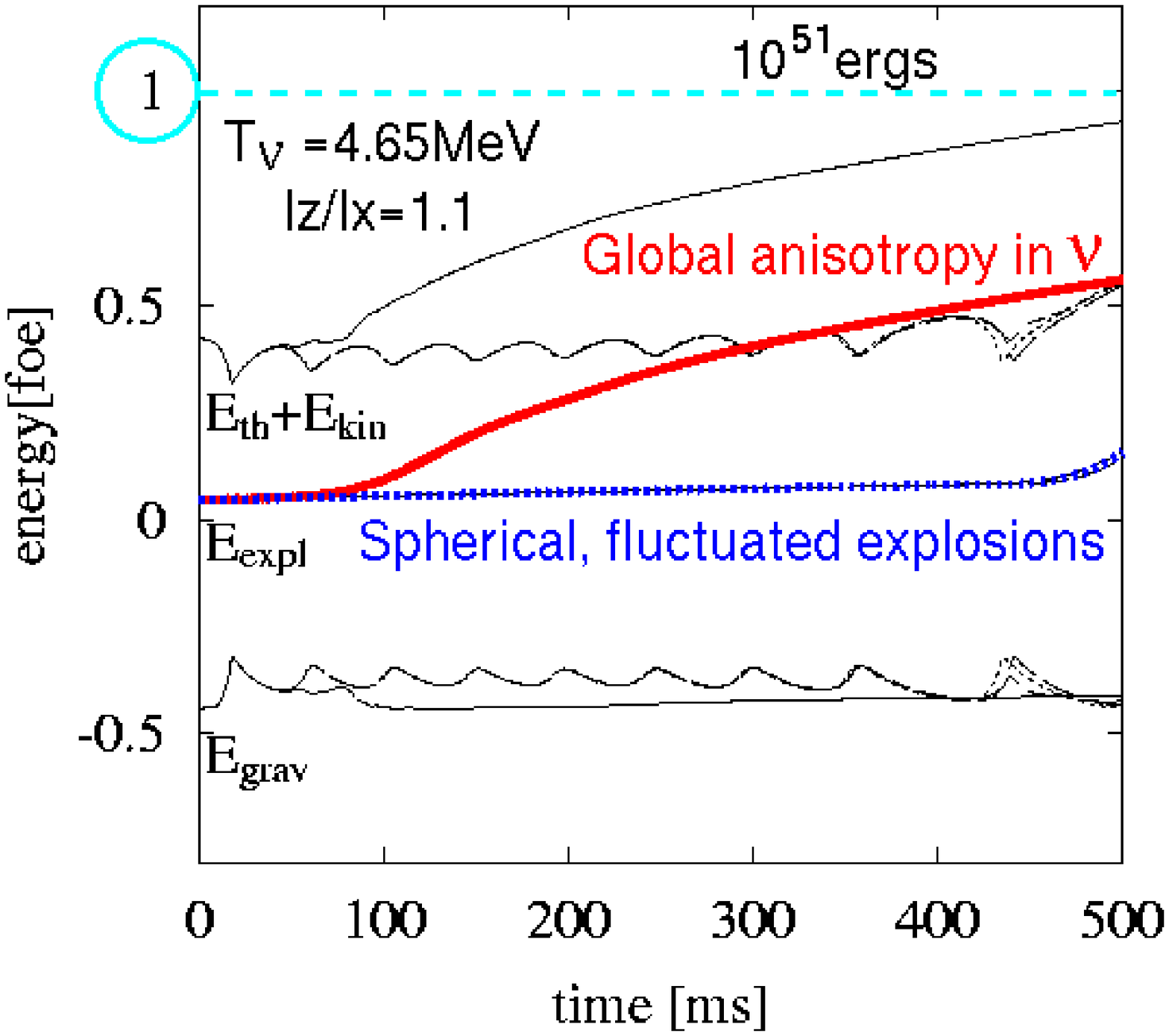}
\end{center}
\caption{Evolution of thermal and kinetic energy (E$_{\rm th}$+E$_{\rm kin}$), 
gravitational energy
(E$_{\rm grav}$), and explosion energy (E$_{\rm expl}$) for the anisotropic 
model of $[l_{z}/l_{x}]_{obs}$ = 1.1 and $T_\nu$ = 4.65 MeV, and
spherical and fluctuated neutrino radiating models (see the text).  
}
\end{figure}

We are going to see that
the explosion energy dramatically increases
as the result of anisotropic neutrino radiation.
Figure~3 shows the evolution of the explosion energy, as well as the thermal, 
kinetic, and gravitational energies.
The thick solid line of the explosion energy ($E_{\rm expl}$) represents
the explosion caused by anisotropic neutrino radiation for the model of
$[l_{z}/l_{x}]_{obs}$ = 1.1 and $T_\nu$ = 4.65 MeV.
The dotted line of $E_{\rm expl}$ 
in Fig.~3 is for the case when the neutrino field is fluctuated in space
(Burrows \etal~1995).
The thin solid line of $E_{\rm expl}$ in Fig.~3, 
which is difficult to distinguish from the
dotted line although, shows the case for the spherical explosion.
We see in Fig.~3 that the simulations fail to explode when the neutrino field is 
spherically symmetric and fluctuated, but may produce an adequate explosion 
when the field is reasonably anisotropic:
The difference between the model with global anisotropy and the other models is extremely
remarkable.
The degree of anisotropy assumed here, $[l_{z}/l_{x}]_{obs}$ = 1.1, 
can be translated into the rotational period of 
$\sim$ 15 ms on the assumption of the Maclaurin 
spheroid, and hence this is considered to be in the reasonable range 
as the rotational period of a proto-neutron star.

In the previous studies
we have performed simulations with various types of models and 
concluded that
the global anisotropy of the neutrino radiation is the
most effective mechanism of increasing the explosion energy 
among spherically symmetric and fluctuated neutrino radiating models
when the total neutrino luminosity is given.
We also found that
the effect of anisotropic neutrino radiation on the explosion 
energy appears with a small value ($[l_{z}/l_{x}]_{obs}$ $\sim$ 1.05) 
and the effect saturates 
with a certain value ($[l_{z}/l_{x}]_{obs}$ $\sim$ 1.2).
We emphasize that a large anisotropy in the neutrino radiation is {\em not}
required for the explosion mechanism. 
We also remark that
Shimizu \etal~(2001) reported that the anisotropy of 
$[l_{z}/l_{x}]_{obs}$ = 1.2 in the neutrino radiation is equivalent to a convective
model with an initial perturbation 
of 10\% in the density of the infalling material,
as far as the effect on the explosion energy is considered. 

\section{Reasoning of Powerful Explosion Induced by Anisotropic $\nu$ Radiation}

We present in this section 
the reasoning of the powerful explosion caused by 
the globally anisotropic neutrino radiation.
The crucial point here is that
the balance between neutrino heating and cooling.
This determines the 
evolution of the explosion energy.
Here the neutrino heating is caused by neutrino absorption due to nuclei and
neutrino scattering off electrons and positrons,
and hence the heating rate depends on the {\em neutrino} temperature, $T_\nu$: 
The heating rate varies as $T_\nu^6$. 
On the other hand, the neutrino cooling is caused by 
neutrino emission due to electron captures and also 
by thermal (photo, pair, plasma) neutrino emission processes.  
The cooling rate is then a steep function
of the {\em matter} temperature $T$, roughly proportional to $T^6$.

\begin{figure}[t]
\begin{center}
\includegraphics[scale=0.4]{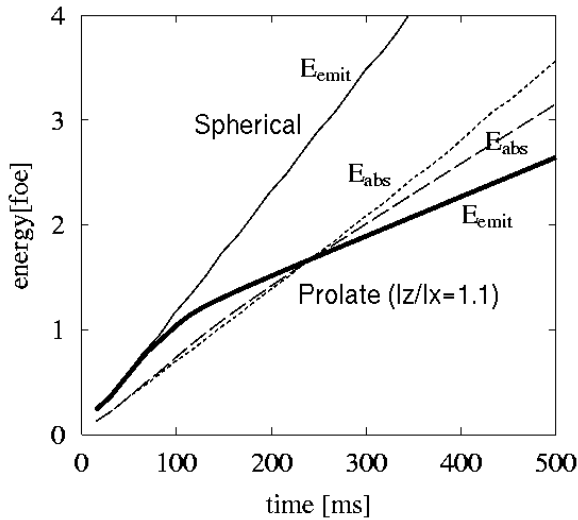}
\end{center}
\caption{
Accumulation of the absorbed and emitted energies due to neutrino 
heating ($E_{\rm abs}$) and cooling ($E_{\rm emit}$).
The thick solid line and the dashed line represent, respectively, 
$E_{\rm emit}$ and $E_{\rm abs}$
for the anisotropic model of $[l_{z}/l_{x}]_{obs}$ = 1.1.
The thin solid line and the dotted line represent 
$E_{\rm emit}$ and $E_{\rm abs}$, respectively, for 
the spherically symmetric model.
For both models neutrino temperature $T_\nu$ = 4.65 MeV is adopted.
Note the significant difference in the emitted energy.
}
\end{figure}

Figure~4 shows the accumulation of the absorbed and emitted energies due to 
neutrino heating and cooling for
the globally anisotropic and for the spherical models.
As mentioned before, 
the locally intense neutrino radiation in the anisotropic model results in local
heating along the polar axis.
The thermal pressure around the heated matter becomes high, and hence
it pushes a part of the shock wave outward.
Namely, the shock revival occurs 
in one direction first along the polar axis.
Then the pressure gradient does work along the shock front,
and the shock revival prevails into all directions
(note the Rankine-Hugoniot condition).
This leads to an earlier shock revival than the spherical model.
As the shock expands, the {\em matter} temperature behind the shock decreases rapidly.
The neutrino cooling rate, which is very sensitive to the matter temperature, 
accordingly drops rapidly.
On the other hand, the neutrino heating rate remains almost unchanged between 
the the anisotropic and the spherical models.
Heating dominates cooling as a result.  
These are seen in Fig.~4.

Eventually, the explosion energy increases because of the suppression of 
neutrino cooling. We note that
the position where the cooling rate suddenly starts to decrease
in Fig.~4
({\em i.e.}, t~$\sim$~90 ms for the anisotropic model)
is equivalent to the time
when the explosion energy suddenly starts to rise up in Fig.~3.
This time corresponds to the shock revival time.

\section{Concluding Remarks: Problems and the Future}

In this article it has been demonstrated that
the explosion energy in the globally anisotropic neutrino radiation model
increases most effectively among 
spherically symmetric and fluctuated radiating models 
when the total neutrino luminosity is given.
As a matter of fact, the expanding materials of 
SN~1987A suggestively has a prolate geometry (Wang \etal~2002). 
We conclude that anisotropic neutrino radiation or locally intense
neutrino heating can be alternative mechanism for a ``successful'' explosion
other than the so far suggested convective trigger.
 
The enhancement of total neutrino luminosity itself is actually 
the most important ingredient
in the delayed explosion mechanism for driving an energetic explosion 
as concluded by Janka \& M\"uller (1995, 1996).
If the total neutrino luminosity adopted in the simulations is sufficiently large, 
any models may easily explode regardless of the type of the explosion.
However, such treatment often leads to the problem of Ni overproduction,
especially in the case of essentially spherical models.
Because of this, the total neutrino luminosity cannot be simply increased.
Our results show that the 
enhancement of the total neutrino luminosity is not always necessary.
Self-consistent simulations of the rotating supernova core and
anisotropic neutrino radiation from the beginning of the gravitational 
collapse are now under progress.

Since we do not know at present whether the neutrino luminosity in a real
supernova explosion is strong enough or not, or whether it has been 
reproduced correctly in the past simulations, we have to wait for the answer
to whether anisotropy in the neutrino radiation field is generally 
important or not, until the neutrino opacity and the numerical technique
for the neutrino transport are improved in future calculations.
Further studies on nuclear equation of state, inelastic neutrino
scattering, and weak interactions on nuclei at subnuclear densities
and at finite temperatures are indeed required.   
Also, future detections of the neutrino spectra by, for example, 
Hyper-Kamiokande experiment and MOON experiment (e.g., Nomachi \etal~2003)
will surely give crucial information on the mechanism of core-collapse supernovae.
Signature of anisotropy in the neutrino radiation field may be 
caught in these future facilities together with future gravitational
wave detections from supernovae.


\end{document}